\documentclass[a4paper,twocolumn,11pt,unpublished]{quantumarticle}

\pdfoutput=1

\makeatother

\usepackage{amsmath} 
\usepackage{amsfonts}
\usepackage{graphicx}
\usepackage{enumitem}
\usepackage{listings}
\usepackage{color}
\usepackage{subcaption}
\usepackage{hyperref}

\usepackage[
  style=numeric-comp,
  sorting=none,
  backend=biber,
  doi=false,
  url=false,
  hyperref=auto,
  giveninits=true,
  uniquename=false,
  maxbibnames=6,
  minbibnames=6
]{biblatex}
\DeclareFieldFormat[article]{title}{#1}
\DeclareFieldFormat[inproceedings]{title}{#1}
\DeclareFieldFormat[incollection]{title}{#1}
\renewbibmacro{in:}{}

\DeclareFieldFormat{pages}{#1}
\DeclareFieldFormat{pagetotal}{#1}

\renewbibmacro*{journal+issuetitle}{%
  \usebibmacro{journal}%
  \setunit*{\addspace}%
  \usebibmacro{volume+number+eid}%
  \setunit{\addcomma\space}%
  \newunit}

\renewbibmacro*{volume+number+eid}{%
  \printfield{volume}%
  \setunit*{\addnbthinspace}
  \printfield{eid}}

\renewbibmacro*{issue+date}{%
  \newunit}

\renewbibmacro*{doi+eprint+url}{%
  \iftoggle{bbx:doi}
    {\printfield{doi}}
    {}%
  \newunit\newblock
  \iftoggle{bbx:eprint}
    {\usebibmacro{eprint}}
    {}%
  \newunit\newblock
  \iftoggle{bbx:url}
    {\usebibmacro{url+urldate}}
    {}%
  \setunit{\addspace}%
  \printtext[parens]{\printfield{year}}
}

\renewbibmacro*{title}{%
  \ifboolexpr{
    test {\iffieldundef{url}}
    and
    test {\iffieldundef{doi}}
  }
  {
    \printtext[title]{%
      \printfield[titlecase]{title}%
      \setunit{\subtitlepunct}%
      \printfield[titlecase]{subtitle}%
    }%
  }
  {
    \iffieldundef{doi}
      {
        \printtext[title]{%
          \href{\thefield{url}}{%
            \printfield[titlecase]{title}%
            \setunit{\subtitlepunct}%
            \printfield[titlecase]{subtitle}%
          }%
        }%
      }
      {
        \printtext[title]{%
          \href{https://doi.org/\thefield{doi}}{%
            \printfield[titlecase]{title}%
            \setunit{\subtitlepunct}%
            \printfield[titlecase]{subtitle}%
          }%
        }%
      }%
  }%
}

\hypersetup{
    hidelinks,
    colorlinks=true,
    breaklinks=true,
    citecolor=MediumPurple,
    filecolor=LimeGreen,
    linkcolor=MediumBlue,
    urlcolor=MediumPurple
}
\usepackage[svgnames]{xcolor}

\newcommand{\ket}[1]{|#1\rangle}

\newcommand{\pytheus}{\textsc{PyTheus }}
\newcommand{\PyTheus}{\textsc{PyTheus}}

\newcommand{\AIMandel}{\textnormal{\textsc{AI-Mandel }}}

\newcommand{\mpl}{Max Planck Institute for the Science of Light, Erlangen, Germany}
\newcommand{\jena}{Institut für Festkörpertheorie und Optik, Friedrich-Schiller-Universität Jena, Jena, Germany}
\newcommand{\tuebingen}{Machine Learning in Science Cluster, Department of Computer Science, Faculty of Science, University of Tübingen, Germany}

\begin{document}

\title{Automated discovery of high-dimensional multipartite entanglement with photons that never interacted}

\author{Sören Arlt}
\email{soeren.arlt@uni-tuebingen.de}
\affiliation{\tuebingen}
\affiliation{\mpl}

\author{Mario Krenn}
\email{mario.krenn@uni-tuebingen.de}
\affiliation{\tuebingen}

\author{Xuemei Gu}
\email{xuemei.gu@uni-jena.de}
\affiliation{\jena}

\begin{abstract}
Quantum entanglement across spatially separated network nodes is conventionally established through the distribution of photons from a common source or via entanglement swapping that relies on Bell-state measurements and pre-shared entanglement. Path identity, where the emission origins of photons from different sources are made indistinguishable, offers an alternative route. We show that this mechanism enables complex multipartite, high-dimensional, and even logical entanglement between remote nodes whose photons never interacted. Our schemes require neither direct photon interaction, pre-shared entanglement, nor Bell-state measurements, highlighting a distinct resource for distributed quantum communication and computation. All of the solutions were discovered automatically using highly efficient computational design tools, indicating the potential for scientific inspiration from computational algorithms.
\end{abstract}

\maketitle

\section{Introduction}
Quantum entanglement establishes correlations between distant particles that defy classical description. Apart from its foundational importance, entanglement has become a crucial resource enabling applications in quantum networks, secure communication, distributed computation, and precision sensing.

\begin{figure*}[!t]
    \centering
    \includegraphics[width=1\textwidth]{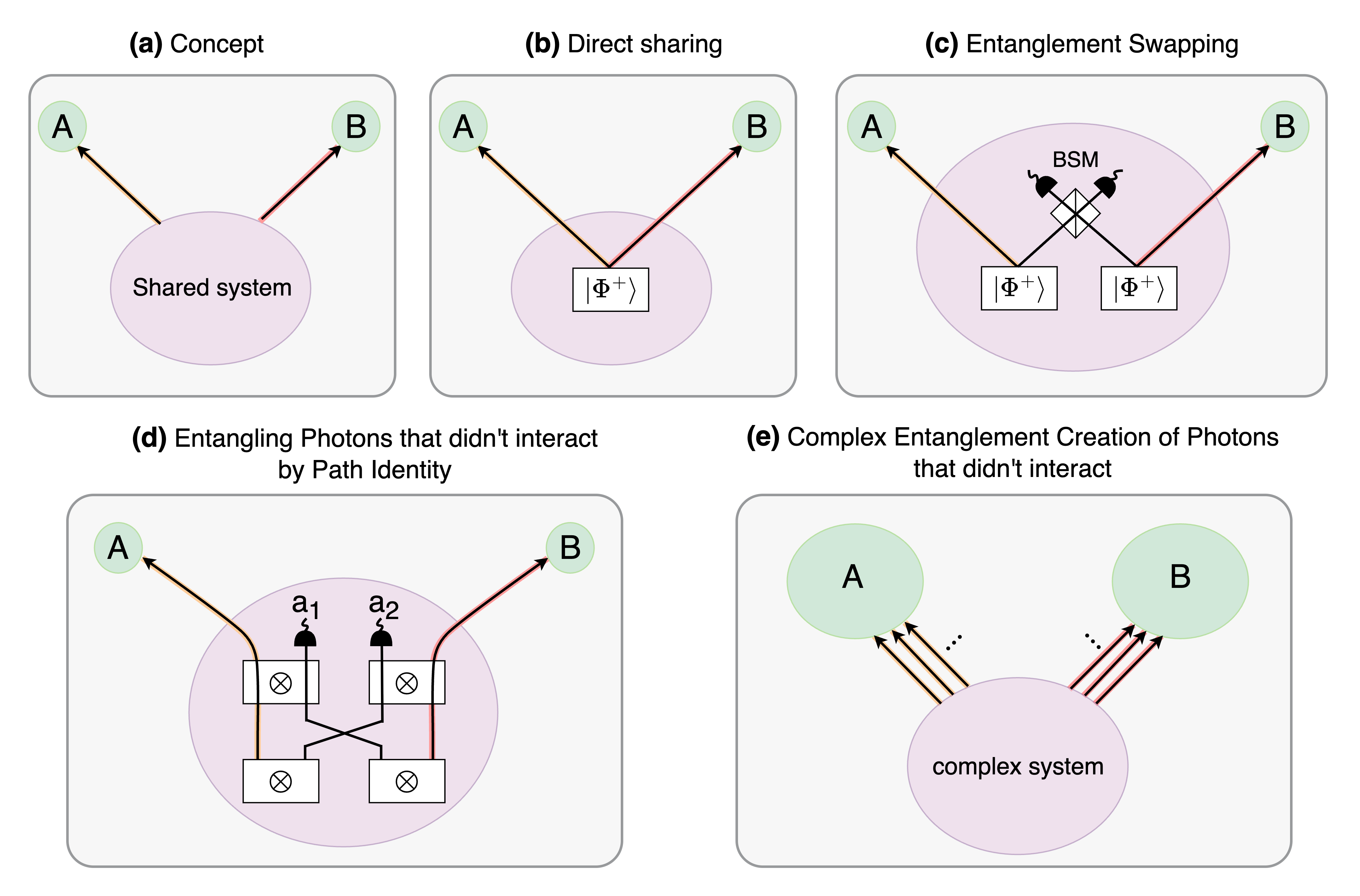}
    \caption{\textbf{Methods for establishing entanglement across separated locations.} \textbf{(a)}. We consider three stations: a shared system and two distant sites A and B. \textbf{(b)}. The simplest way is to generate entanglement at a common source (for example a Bell state $\ket{\Phi^+}$, as indicated in the white rectangle) and then sent to A and B. This scenario was first described by Einstein, Podolsky, and Rosen in 1935 \cite{einstein1935can} and later demonstrated by Freedman and Clauser in 1972 \cite{freedman1972experimental}. \textbf{(c)}. Entanglement swapping allows photons at A and B to become entangled without interacting before. Two pre-shared Bell pairs $\ket{\Phi^+}$ and a Bell state measurement (BSM) at the central node are used \cite{zukowski1993event}, and this was demonstrated in 1998 \cite{pan1998experimental}. \textbf{(d)}. Path identity generates entanglement between photons that never interacted, without Bell-state measurements and pre-shared entanglement \cite{krenn2017entanglement,ruiz2023digital}. This was discovered by \pytheus \cite{ruiz2023digital} and experimentally demonstrated in 2024 \cite{wang2024entangling}. The photon sources produce photon pairs or product states, indicated by the symbol $\otimes$. \textbf{(e)}. In the present work, complex multipartite and high-dimensional entangled states are established across separated locations without direct interaction, pre-shared entanglement, or Bell state measurements.}
    \label{fig:overall}
\end{figure*}

Establishing entanglement between spatially separated locations is a central requirement for the operation of quantum networks. In the conventional approach, this is achieved by allowing particles to interact locally at a single site or a common source and then distributing them to remote stations (Fig.~\ref{fig:overall}(b)). This picture was first discussed in the seminal 1935 work of Einstein, Podolsky, and Rosen, who considered two systems that interact for a finite duration and are subsequently separated without any further interaction, revealing non-local consequences predicted by quantum mechanics \cite{einstein1935can}. However, entanglement does not always have to originate from direct interaction.


Entanglement swapping, proposed in 1993 \cite{zukowski1993event}, allows two previously independent particles to become entangled via a Bell-state measurement, even though they never interacted (see Fig.~\ref{fig:overall}(c)). This idea was experimentally demonstrated in 1997 \cite{pan1998experimental} and has since become a cornerstone of quantum information processing, enabling applications such as quantum repeaters for long-distance quantum communications \cite{duan2001long, azuma2023quantum}.

Recently, a fundamentally different approach was introduced that requires neither direct interaction, pre-established entanglement, nor Bell-state measurements \cite{ruiz2023digital, wang2024entangling} (Fig.~\ref{fig:overall}(d)). It is based on \textit{path identity}, where the origins of photons are created such that one cannot tell which source created which photon \cite{krenn2017entanglement}. Recent experimental demonstrations verify that these new concepts can be implemented both in integrated photonics platforms \cite{feng2023chip,bao2023very} and in bulk-optics \cite{qian2023multiphoton,hu2025observation}. The underlying indistinguishability enables entanglement between independent particles that never interacted before.
The idea was first discovered by \PyTheus, a computational tool for automatically designing quantum-optics experiments \cite{ruiz2023digital}, and was experimentally demonstrated in 2024 for creating Bell states \cite{wang2024entangling}.

Here we extend this idea dramatically, showing that path identity can generate complex forms of entanglement across nodes whose photons never met (Fig.~\ref{fig:overall}(e)). Specifically, we propose experimental schemes for establishing multi-photon GHZ and W states to high-dimensional entanglement and encoded logical entanglement, including hybrid logical–physical variants. Creating logical entanglement in this way could become important for distributed quantum computers in large future quantum networks \cite{kimble2008quantum}. None of these schemes require direct interaction, pre-shared entanglement, or Bell-state measurements. All the configurations were discovered by \pytheus \cite{ruiz2023digital}, demonstrating how computational tools can reveal nonintuitive quantum-optical designs \cite{krenn2020computer,krenn2023artificial}. Together, these results open new opportunities for distributed quantum information processing in quantum networks.

\begin{figure*}[!t]
    \centering
    \includegraphics[width = 0.85\textwidth]{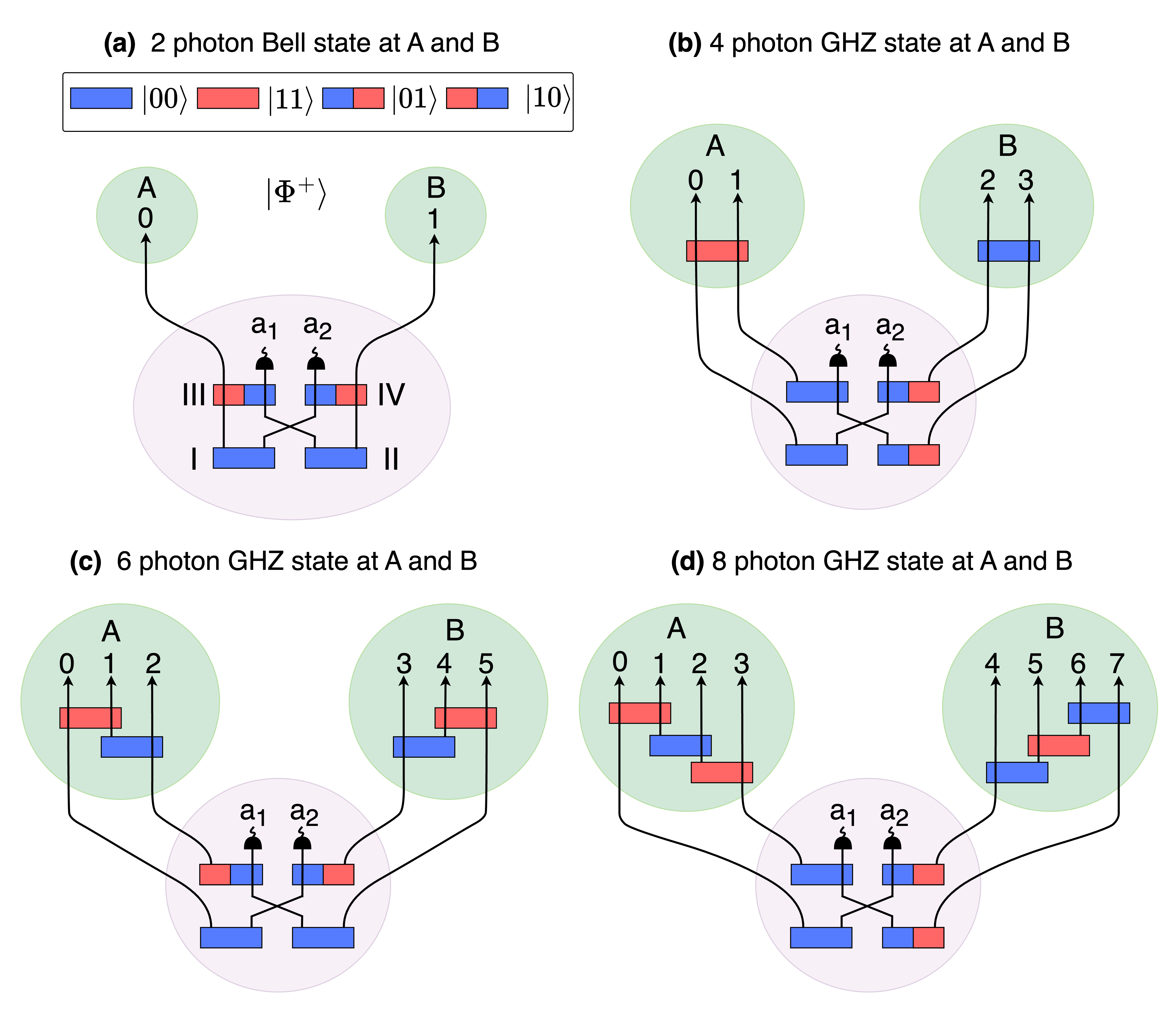}
    \caption{\textbf{Multi-photon GHZ states with photons that never interacted at two separated locations.} \textbf{(a)} The simplest case: a two-photon Bell state, which can be considered as the two-photon instance of a GHZ state. The remote photons in paths 0 and 1 are at locations $A$ and $B$, while ancillary photons are detected in paths $a_{1},a_{2}$. \textbf{(b–d)} Extension to four-, six-, and eight-photon GHZ states. Neither pre-shared entanglement nor Bell-state measurements are required. Photon pairs are generated in nonlinear crystals (for example, the blue-blue rectangle denotes a photon pair in the state $\ket{00}$). Photons at sites $A$ and $B$ become entangled without ever interacting.}
    \label{fig:GHZ}
\end{figure*}

\begin{figure*}[!t]
    \centering
    \includegraphics[width =1\textwidth]{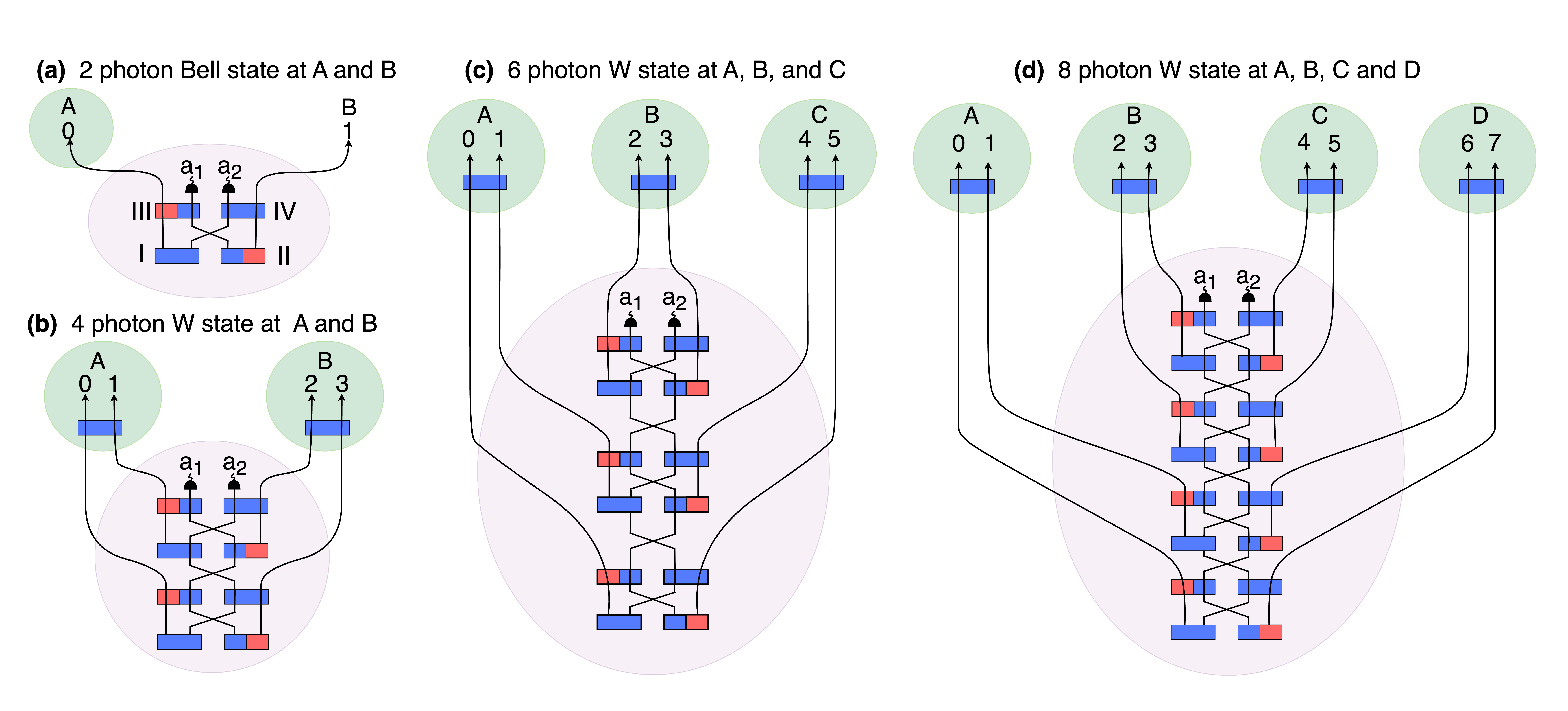}
    \caption{\textbf{Multi-photon W states with photons that never interacted at separated locations.}  \textbf{(a)} A two-photon Bell state $\ket{\Psi^+}$, can be considered as the two-photon case of a W state. \textbf{(b–d)} Four-, six-, and eight-photon W states distributed across two, three, and four locations. Photons at different sites that never met become entangled.}
    \label{fig:Wstate}
\end{figure*}

\subsection{Idea generation via \AIMandel}
The initial research directions that led to this work were first conceived by \AIMandel \cite{arlt2025towards}, an LLM-agent system for autonomous idea generation in quantum physics that is coupled to \pytheus. In our workflow, \AIMandel proposes ideas, iteratively screens them for novelty and feasibility using dedicated agent roles, and then translates accepted ideas into concrete, automatically discovered photonic designs via the tool interface. The human authors subsequently interpreted the machine-generated designs, validated the underlying mechanisms, and generalized them into the broader principles and families of schemes presented in this article. 

\section{Results}
\subsection{Greenberger–Horne–Zeilinger states}
Multipartite entanglement gives rise to phenomena that do not appear in the bipartite case \cite{lawrence2014rotational, erhard2020advances} and enables forms of nonlocality that cannot be reduced to pairwise correlations \cite{coiteux2021no}. A canonical example is the Greenberger–Horne–Zeilinger (GHZ) state \cite{greenberger1989going}, defined as
\begin{align}
    \ket{GHZ}_n=\frac{\ket{0}^{\otimes n}+\ket{1}^{\otimes n}}{\sqrt{2}}.
\end{align}

We used \pytheus to discover configurations that generate GHZ entanglement across separated locations, without any direct photon interaction, pre-established entanglement or Bell-state measurements.

Fig.~\ref{fig:GHZ}(a) illustrates the bipartite case, which corresponds to the $n = 2$ instance of a GHZ state and is conceptually similar to Fig.~\ref{fig:overall}(d). In this simple setup, all four nonlinear crystals are pumped coherently, and the pump power is chosen such that only two photon pairs are produced simultaneously, while higher-order emission events can be neglected \cite{krenn2017entanglement}. The colors (red and blue) indicate the degree of freedom in which the photons are encoded. Sources I and II each generate photon pairs in the state $\ket{00}$ (blue–blue), while source III produces $\ket{10}$ (red-blue) and source IV produces $\ket{01}$ (blue-red). The mode number can represent polarization, orbital angular momentum, or another photonic degree of freedom. The photons from sources I and II are aligned with those from sources III and IV such that their paths are overlapped. In the experiment, n-fold coincidence detection is used to obtain the quantum state, with one photon detected in each output path. This can happen if the two photon pairs originate either from sources I and II or from sources III and IV. The coherent superposition of these two cases yields the normalized state 
\begin{align}
\ket{\psi}=\frac{(\ket{00}_{AB} + \ket{11}_{AB})\ket{00}_{a_1 a_2}}{\sqrt{2}},
\end{align}
which gives the Bell state $\ket{\Phi^+}_{AB}$ between the photons that never interacted at locations A and B. Interestingly, unlike entanglement swapping, these setups require neither pre-shared entangled pairs nor Bell-state measurements.

Fig.~\ref{fig:GHZ}(b)–(d) extend this setup to four-, six-, and eight-photon GHZ states distributed across two distant sites. In all of these setups, entanglement is generated between photons that never met at the two locations. Theses experimental schemes were discovered by \pytheus based on a graph representation of quantum experiments \cite{krenn2017quantum, gu2019quantum,gu2019graph3,krenn2021conceptual}, where each graph can be translated into quantum optical setups. The graph solutions for the setups in Fig.~\ref{fig:GHZ} are provided in the Appendix.

\subsection{W states}
Alongside the GHZ state, a second, fundamentally different class of multipartite entanglement is the W state, inequivalent to GHZ states under local operations and classical communication \cite{dur2000three}. A key feature of W states is their inherent robustness against particle loss. This makes W states particularly valuable for distributed quantum communication, entanglement routing, and fault-tolerant quantum networks.

For $n$ particles, the W state is a symmetric superposition in which exactly one particle is in state $\ket{1}$ and the others are in $\ket{0}$, given by
\begin{align}
    \ket{W}_n=\frac{\ket{100...0}+\ket{010...0}+...+\ket{00...01}}{\sqrt{n}}.
\end{align}
Due to their robustness and broad applicability, W states have been generated through several approaches, including the use of pre-shared entanglement \cite{eibl2004experimental, tashima2009local} and entanglement-distribution methods such as quantum repeaters \cite{miguel2023quantum}. In contrast to these conventional schemes, we use \pytheus to design configurations that generate W states across spatially separated locations without direct interaction, pre-shared entanglements or Bell-state measurements. The photons originate from different nonlinear crystals, and the coherent superposition of their indistinguishable emission processes leads to the W state.

Fig.~\ref{fig:Wstate} shows W state generation across multiple locations. The simplest case in Fig.~\ref{fig:Wstate}(a) shows a Bell state $\ket{\Psi^+}$ between two locations, which corresponds to the $n = 2$ instance of a W state. Fig.~\ref{fig:Wstate}(b)–(d) present increasingly complex examples: four-, six-, and eight-photon W states at two, three, and four remote nodes, respectively. Each configuration realizes entanglement among spatially separated photons without requiring them to share a common origin, pre-established entanglement, or Bell-state measurements. These experimental schemes were discovered from \pytheus and the graph solutions for the experiments are shown in the Appendix.

\begin{figure}[!t]
    \centering
    \includegraphics[width = 0.45\textwidth]{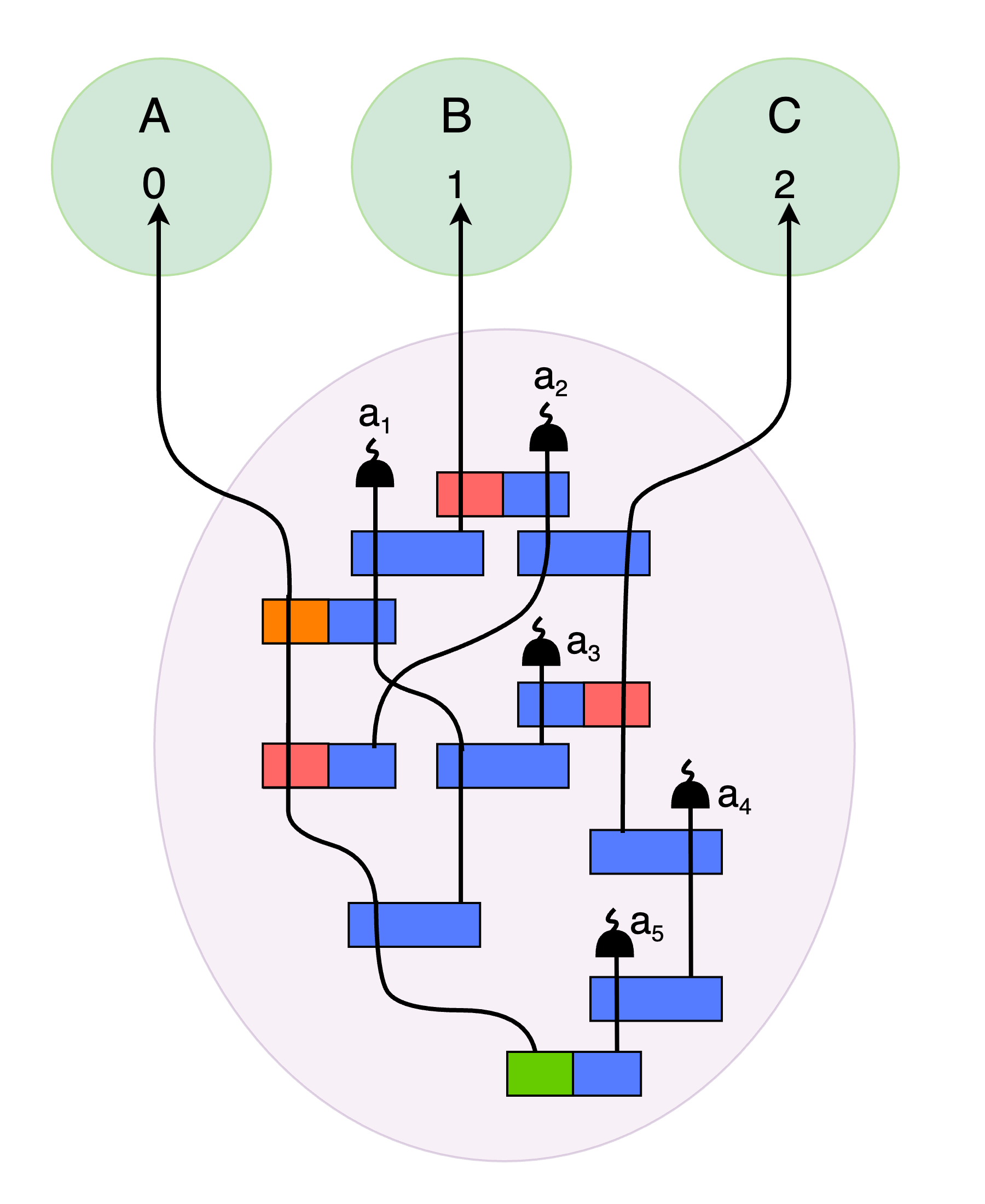}
    \caption{\textbf{SRV(4,2,2) state distributed across three spatially separated locations with photons in paths 0--2 that never interacted.} 
    Basis states are color-coded: $\ket{3}$ (yellow), $\ket{2}$ (green), $\ket{1}$ (red), and $\ket{0}$ (blue). Ancillary photons in paths $a_{1}$--$a_{5}$ are prepared in $\ket{0}$.}
    \label{fig:SRV422}
\end{figure}
\subsection{Schmidt-rank vector states}
Multiparticle entangled states in high dimensions exhibit a much richer internal structure than qubit-based systems. One way to characterize this structure is through the \emph{Schmidt-rank vector} (SRV), which specifies the dimensionality of each subsystem by the rank of its reduced density operator \cite{huber2013structure, huber2013entropy}. For instance, an SRV$(d_A,d_B,d_C)$ state is one where the reduced density matrices of the three subsystems $A$, $B$, and $C$ have Schmidt ranks $d_A$, $d_B$, and $d_C$, respectively. 
Such states give rise to phenomena that only emerge when both the number of particles and dimensions exceed two. They have been realized experimentally \cite{erhard2018experimental, malik2016multi, hu2020experimental} and discovered through computer-designed setups \cite{krenn2016automated, ruiz2023digital}. They are especially relevant for layered quantum communication \cite{pivoluska2018layered} and quantum networks, where nodes may operate with different dimensional capabilities.

As an example, we consider a tripartite system and generate an SRV(4,2,2) state across three spatially separated locations without any direct interaction, pre-established entanglement, or Bell-state measurements:
\begin{align}
    \ket{\psi}_{ABC}=\frac{\ket{311}+\ket{210}+\ket{101}+\ket{000}}{2}
\end{align}
with reduced density operator ranks
\begin{align}
    \text{rank}(\rho_{A})=4, \,\, \text{rank}(\rho_{B})=2, \,\, \text{rank}(\rho_{C})=2, \nonumber
\end{align}
where the photon at location A occupies a four-dimensional subspace, while photons at B and C each lives in two-dimensional subspaces. 

Crucially, the photons at the three locations never interacted and do not originate from a common entangled source. The corresponding experimental design is shown in Fig.~\ref{fig:SRV422}, and its graph solution is provided in the Appendix.

\subsection{Logical qubit and their entanglement} 
Quantum error correction encodes information across multiple \textit{physical} qubits, forming a \textit{logical} qubit that is resilient to local errors, a crucial step towards fault-tolerant quantum information. The encoding of logical qubits enables the identification and, for certain codes, correction of local errors, such as bit flips or phase flips, without destroying the encoded information \cite{gottesman2002introduction, roffe2019quantum}. More precisely, the physical basis states $\ket{0}$ and $\ket{1}$ are mapped to logical states $\ket{0_L}$ and $\ket{1_L}$, whose specific form depends on the chosen error-correcting code \cite{ErrorCorrectionZoo}.

Entangling logical qubits across distant nodes offers a promising route toward fault-tolerant quantum communication and distributed quantum computing \cite{gottesman2009introduction, muralidharan2014ultrafast, nielsen2010quantum}. The simplest example of such encoded entanglement is the \emph{logical Bell state},
\begin{align}
\ket{\Phi^+_L}=\frac{1}{\sqrt{2}}\Big( \ket{0_L,0_L} + \ket{1_L,1_L} \Big),
\label{eq:logicBell}
\end{align}
which can be generated between the photons comprising the two logical blocks, without employing Bell-state measurements, entanglement swapping, or any joint interaction between locations $A$ and $B$. In the following, we explore examples of quantum error-correcting codes that generate the states in Eq.~\eqref{eq:logicBell}. We show logical entanglement can be established between distant locations with particles that never met, without any pre-shared entanglement, direct interaction, or Bell-state measurements.

\subsubsection{Three-qubit repetition code}
The simplest quantum error-correcting code is the three-qubit repetition code, which maps a single-qubit state onto three physical qubits. Depending on the variant, it can correct a single bit-flip or phase-flip error on any of the three qubits. Such codes have been experimentally demonstrated in superconducting qubit systems \cite{reed2012realization, kelly2015state} and trapped ions \cite{schindler2011experimental}. Here we consider the bit-flip version, with the logical encoding
\begin{align}
\ket{0_L}=\ket{000}, \,\,\,\ket{1_L}=\ket{111}.
\end{align}

The corresponding logical Bell state of Eq.~\eqref{eq:logicBell} thus becomes a six-qubit GHZ state \cite{greenberger1989going},
\begin{align}
\ket{\psi} = \frac{1}{\sqrt{2}}\Big(\ket{000}_A\ket{000}_B + \ket{111}_A\ket{111}_B\Big),
\end{align}

An implementation of this encoded entanglement is shown in Fig.~\ref{fig:GHZ}(c). The experiment employs eight coherently pumped photon-pair sources. Each successful event probabilistically emits four indistinguishable photon pairs, generating eight photons in total: six encoding the two logical qubits (paths 0–2 and 3–5) and two ancillary photons detected at ports $a_{1}$ and $a_{2}$. Notably, photons from blocks $A$ and $B$ neither directly interact nor share a common source. No Bell-state measurements are performed, and all sources emit separable photon pairs.

\subsubsection{Quantum-Error-Detection code $[[4,1,2]]$} 
Surface codes are a prominent class of quantum error-correcting codes that protect quantum information through local stabilizer measurements on a lattice of physical qubits \cite{nielsen2010quantum}. Their topological construction provides robustness against local noise and supports scalable fault-tolerant architectures \cite{kitaev2003fault, dennis2002topological, fowler2012surface}. Surface codes are a well-known family of geometrically local CSS codes that protect quantum information through stabilizer measurements on a lattice.
As a minimal illustration of such CSS-type constructions, we consider here the four-qubit [[4,1,2]] code\cite{ErrorCorrectionZoo}, where the notation $[[n,k,d]]$ indicates $n=4$ physical qubits encoding $k=1$ logical qubit with code distance $d=2$. We consider the following logical basis states \cite{leung1997approximate}
\begin{align}
\ket{0_L} = \frac{1}{\sqrt{2}}\left(\ket{0000}+\ket{1111}\right),\label{eq:surfacecoding0}\\
\ket{1_L} = \frac{1}{\sqrt{2}}\left(\ket{0011}+\ket{1100}\right),
\label{eq:surfacecoding1}
\end{align}
which are stabilized by the operators $X^{\otimes4}$ and $Z^{\otimes4}$, where $X$ and $Z$ denote the Pauli operators. These stabilizers allow single-qubit Pauli errors to be detected by monitoring changes in their eigenvalues, as any such error flips the eigenvalue of either $X^{\otimes4}$ or $Z^{\otimes4}$.

\begin{figure}
    \centering
    \includegraphics[width = 0.5\textwidth]{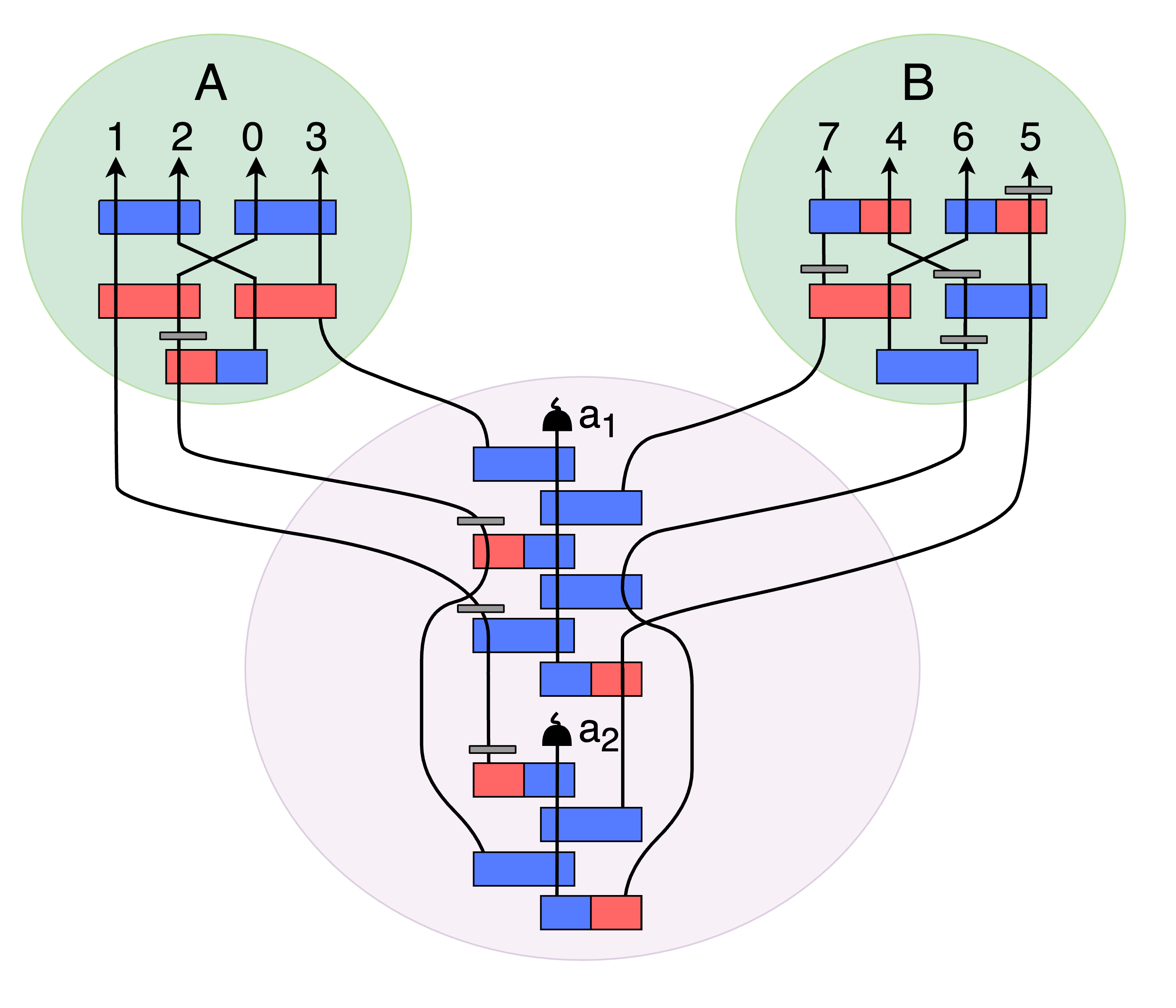}
    \caption{\textbf{Logical Bell state with the error detection code $[[4,1,2]]$ encoding.} Each logical qubit is encoded into four physical qubits (photons), located at sites $A$ (paths 0–3) and $B$ (paths 4–7). The logical Bell state is established between the two sites even though photons at $A$ and $B$ never met. No direct interaction, pre-shared entanglement, or Bell-state measurements are required.}
    \label{fig:surface_code}
\end{figure}
Substituting Eq.~\eqref{eq:surfacecoding0}-\eqref{eq:surfacecoding1} into Eq.~\eqref{eq:logicBell} gives the logical Bell state in the form of 
\begin{align}
\nonumber
\ket{\psi}=&\frac{1}{2\sqrt{2}}\Big(\ket{00000000}+\ket{00001111}\\\nonumber
&\quad+\ket{11110000}+\ket{11111111}\\\nonumber
&\quad+\ket{00110011}+\ket{00111100}\\
&\quad+\ket{11000011}+\ket{11001100}
\Big).
\end{align}

The setup discovered by \pytheus is shown in Fig.~\ref{fig:surface_code}. There, logical qubit (photons in paths 0–3) at location $A$ never interact with the logical qubit (photons in paths 4–7) at location $B$, requiring no direct interaction, pre-established entanglement, or Bell-state measurements.

\subsection{High-dimensional logical entanglement}

So far, we have discussed encoding schemes based on qubits, which are inherently two-dimensional systems. In recent years, high-dimensional quantum systems have reached a level of experimental maturity in photonic \cite{erhard2020advances}, superconducting \cite{cervera2022experimental}, and trapped-ion platforms \cite{ringbauer2022universal}, enabling the practical exploration of multi-level encodings. Encoding quantum information in higher-dimensional systems could provide improved code parameters \cite{gottesman1998fault, knill1996non}.

We show a form of hybrid logical–physical entanglement, where an encoded logical qudit at one site is entangled with a single physical qudit at another. An example is given by
\begin{align}
\ket{\psi}=\frac{1}{\sqrt{3}}\Big( \ket{0_L,0} + \ket{1_L,1} + \ket{2_L,2}\Big),
\label{eq:logic3D}
\end{align}
which is a high-dimensional Bell-type state between a logical system and a physical carrier. Such systems might become relevant when entanglement between distant quantum computers are established in complex quantum networks. In the following, we explore examples of quantum error-correcting codes that generate the states in Eq.~\eqref{eq:logic3D} between distant locations with particles that never met.

\subsubsection{Three-qutrit code $[[3,1,2]]_3$}
An interesting example is the three-qutrit code $[[3,1,2]]_3$, which is a prime-qudit Calderbank–Shor–Steane stabilizer code that encodes one logical qutrit into three physical qutrits \cite{ErrorCorrectionZoo, cleve1999share}. This code can detect single-qutrit errors and saturate the quantum Singleton bound, making it a smallest nontrivial quantum maximum-distance separable code. Its logical basis states are \cite{ErrorCorrectionZoo}
\begin{align}
\ket{0_L}=\frac{1}{\sqrt{3}}\bigl(\ket{000}+\ket{111}+\ket{222}\bigr),\label{eq:css0}\\
\ket{1_L}=\frac{1}{\sqrt{3}}\bigl(\ket{012}+\ket{120}+\ket{201}\bigr), \label{eq:css2}\\
\ket{2_L}=\frac{1}{\sqrt{3}}\bigl(\ket{021}+\ket{102}+\ket{210}\bigr).
\label{eq:css2}
\end{align}
Each logical state exhibits a cyclic permutation structure within its superposition. Beyond its error-correcting capability, this code also realizes a quantum secret-sharing scheme \cite{cleve1999share} and can be used as a minimal model for the AdS/CFT holographic duality \cite{almheiri2015bulk}.

We now want to establish the entanglement as described in Eq.~\eqref{eq:logic3D}. Taking the logical encodings from Eq.~\eqref{eq:css0}-\eqref{eq:css2}, we can obtain the final quantum state of Eq.~\eqref{eq:logic3D} as
\begin{align}
\nonumber\ket{\psi}_{A,B}&=\frac{1}{3}\Big(\ket{0000} +\ket{1110} +\ket{2220} \\ \nonumber
&\quad + \ket{0121} + \ket{1201}+ \ket{2011} \\ 
& \quad + \ket{0212}+ \ket{1022} + \ket{2102}\Big).
\end{align}
where the first three qutrits correspond to the logical subsystem at $A$, and the last qutrit to the physical carrier at $B$. This state establishes entanglement between the logical and physical qutrits even though they never directly interact, offering a pathway toward distributed and fault-tolerant quantum communication. The explicit solution was obtained by \PyTheus, which is provided in the Appendix.

\subsubsection{Amplitude-Damping Code $[[4,1]]_3$}
Conventional stabilizer codes are typically designed to correct local Pauli errors; since Pauli operators span the operator space, this suffices for arbitrary noise by linearity. When the dominant noise process is known, one can design \emph{noise-adapted} codes that provide more efficient protection for that specific channel. A prominent example is amplitude-damping noise, which is directionally biased toward decay. In this setting, tailored codes can outperform Pauli-symmetric designs with comparable or even fewer physical carriers \cite{leung1997approximate,fletcher2008channel}. Extensions of this idea to higher dimensions have led to explicit constructions of amplitude-damping-adapted qudit codes \cite{dutta2025noise}.

We use the notation $[[n,k]]_d$ to denote a code that encodes \(k\) logical \(d\)-level systems into \(n\) physical carriers. As an example, we consider the qutrit amplitude-damping-adapted code $[[4,1]]_3$ with the logical basis \cite{dutta2025noise}
\begin{align}
\ket{0_L}=\frac{1}{\sqrt{3}}\Big(\ket{0000}+\ket{1111}+\ket{2222}\Big)
\label{eq:ad0}\\
\ket{1_L}=\frac{1}{\sqrt{3}}\Big(\ket{0011}+\ket{1122}+\ket{2200}\Big)
\label{eq:ad1}\\
\ket{2_L}=\frac{1}{\sqrt{3}}\Big(\ket{1100}+\ket{2211}+\ket{0022}\Big)
\label{eq:ad2}
\end{align}
Each codeword has balanced support across the basis levels, which helps mitigate the error. The $[[4,1]]_3$ code is part of a broader family of noise-adapted qudit encodings \cite{ErrorCorrectionZoo,dutta2025noise}.

To form the state in Eq.~\eqref{eq:logic3D}, we substitute Eqs.~\eqref{eq:ad0}–\eqref{eq:ad2} and get
\begin{align}
\nonumber \ket{\psi}_{A,B}&=\frac{1}{3}\Big(\ket{00000} +\ket{11110} +\ket{22220} \\ \nonumber
&\quad + \ket{00111} + \ket{11221}+ \ket{22001} \\ 
& \quad + \ket{11002}+ \ket{22112} + \ket{00222}\Big).
\end{align}
where the first four qutrits comprise the logical system at $A$ and the last is the physical carrier at $B$. This state shows hybrid logical–physical entanglement tailored to amplitude-damping noise. The explicit solution was obtained by \PyTheus, which is provided in the Appendix.

\section{Conclusion}
Our results point to a networking paradigm in which coherently pumped, indistinguishable emission processes act as an entangling resource between remote quantum processors. In contrast to BSM-based schemes, path-identity designs can generate multi-party, high-dimensional, and even logical entanglement without pre-shared entanglement or joint measurements, suggesting an alternative route to logical-by-design quantum networks with remote nodes. It demonstrates that this form of indistinguishability is an underexplored resource for quantum networking.


Other approaches exploit indistinguishable particles to generating entanglement, including schemes based on local identical particle counting and classical communication \cite{castellini2019activating, mahdavipour2024generation, sun2020experimental, wang2022remote}, as well as linear optics with specific post-selection coincidence counts \cite{blasiak2019entangling, blasiak2022arbitrary}. It would be interesting to develop a common conceptual framework unifying these works with path-identity designs, potentially expanding the range of discoverable quantum-network experiments.


Future work could explore the setups in high-gain regime -- for example using recently derived analytical expressions of the two-mode squeezer operator in the Fock basis \cite{gu2025analytical} --, where multi-pair contributions may no longer act merely as noise but could enable new classes of entanglement. Exploring this regime, alongside strategies for loss mitigation and error correction, are interesting directions for further investigation. It also remains an open question to identify the full range of states achievable through path-identity constructions.

Our findings, combined with automated design methods \cite{krenn2020computer,ruiz2023digital}, indicate that a broad landscape of unexplored entanglement schemes remains to be discovered. The successful generation of high-dimensional multipartite entanglement across remote locations without traditional resource requirements opens new possibilities for distributed quantum information processing in future quantum networks.

Our present results also illustrate a broader methodological point: with the release of \AIMandel \cite{arlt2025towards}, LLM-agent driven hypothesis generation coupled to automated discovery tools can help systematically explore large design spaces in quantum optics. Further developing such closed-loop pipelines, linking idea generation, constraint-aware synthesis, and experimental feedback, may accelerate the identification of useful photonic primitives for scalable quantum networks.

\section*{Acknowledgments}
MK acknowledges support by the European Research Council (ERC) under the European Union’s Horizon Europe research and innovation programme (ERC-2024-STG, 101165179, ArtDisQ) and from the German Research Foundation DFG (EXC 2064/1, Project 390727645). X.G. acknowledges support from the NOA Collaborative Research Center and the Alexander von Humboldt Foundation.  

\printbibliography

\newpage
\clearpage
\appendix
\onecolumn
\section{Graph representation of quantum experiments in \pytheus}

\begin{figure*}[!h]
    \centering
    \includegraphics[width=1\textwidth]{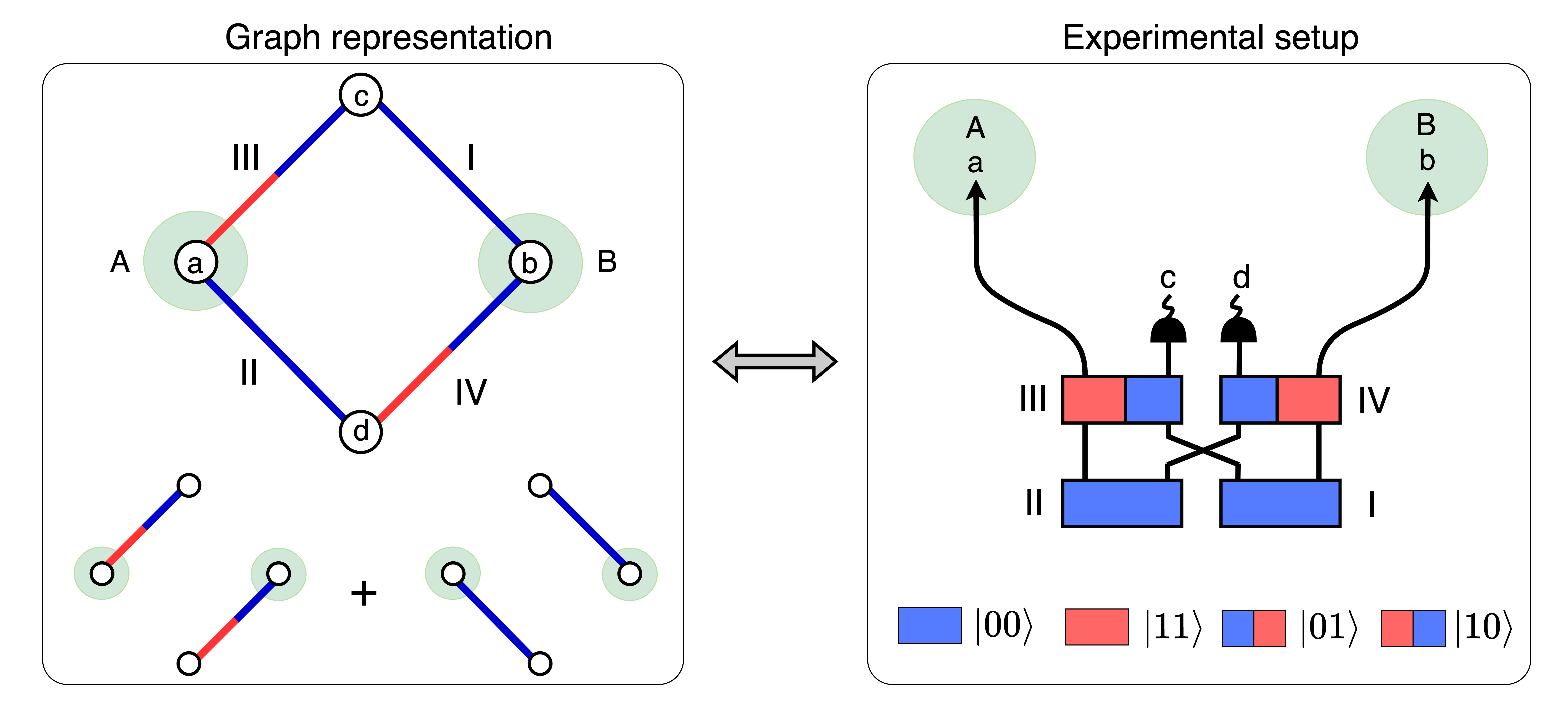}
    \caption{\textbf{Connection between graph theory and quantum optics experiments.} The left panel shows an abstract graph representation of an experiment producing a Bell state with photons that never interacted at locations A and B. The graph has two perfect matchings (subgraphs covering all vertices once), whose coherent superposition yields the state $\ket{\psi} = \omega_{0,a_1}^{1,0}\omega_{1,a_2}^{1,0}\ket{1100} + \omega_{0,a_2}^{0,0}\omega_{1,a_1}^{0,0}\ket{0000}$ (unnormalized). The right panel illustrates the corresponding experimental setup based on the path-identity technique. The squares I–IV denote probabilistic photon-pair sources, color-coded according to the photon's internal degrees of freedom (e.g., polarization, time-bin, or frequency).}
    \label{fig:graph}
\end{figure*}

Our solutions are discovered by \pytheus \cite{ruiz2023digital}, which builds on the connection between quantum optical experiments and graph theory \cite{krenn2017quantum,gu2019quantum,gu2019graph3,krenn2021conceptual}. In this representation, vertices represent photon paths to detectors, edges correspond to correlated photon pairs with complex weights denoting amplitudes, and edge colors encode internal degrees of freedom such as polarization, time-bin, or frequency. This abstract representation can be directly translated into different experimental implementations, including integrated photonics (path encoding) and bulk optics. Figure~\ref{fig:graph} illustrates how a graph maps to a setup using \textit{entanglement by path identity} \cite{krenn2017entanglement}. For further details, we refer to Ref.~\cite{ruiz2023digital}.

The quantum state is generated by the weight function \cite{ruiz2023digital}
\begin{align}
\Phi(\boldsymbol{\omega})=\sum_m \frac{1}{m!} \left( \sum_{e \in E(G)} \omega(e) x^\dagger(e) y^\dagger(e) + \text{h.c.}\right)^m,
\label{eq:weightfunction}
\end{align}
where $E(G)$ is the set of edges, $\omega(e)$ are edge weights with $|\omega(e)|^2<1$, and $\ket{\psi}=\Phi(\boldsymbol{\omega})\ket{\text{vac}}$. The term h.c. stands for hermitian conjugate, which includes annihilation operators. For the graph in Fig.~\ref{fig:graph} with four paths $a,b,c,d$ and two internal modes (0 and 1), this becomes
\begin{align}
\Phi(\boldsymbol{\omega}) \approx \sum_N \frac{1}{N!}(\omega_{a,c}^{1,0}a^\dagger_1 c^\dagger_0
+ \omega_{b,d}^{1,0}b^\dagger_1 d^\dagger_0+\omega_{a,d}^{0,0}a^\dagger_0 d^\dagger_0
+ \omega_{b,c}^{0,0}b^\dagger_0 c^\dagger_0 + \text{h.c.})^N,
\label{eq:pairsource}
\end{align}
where the superscript and subscript on $\omega_{x,y}^{i,j}\in\mathbb{C}$ represent the mode number and optical path, respectively, and $x^\dagger_{k}$ creates a photon in path $x$ with mode $k$. Higher-order pair emissions ($N>2$) in Fig.~\ref{fig:graph} occur with lower probability and usually can be neglected in the low-pump regime.

Experimentally, we condition on $n$-fold coincidence detection, i.e., each path has one photon at each detector. In the graph representation, this corresponds to selecting subsets of edges that cover each of the $n$ vertices exactly once (perfect matchings). For Fig.~\ref{fig:graph}, neglecting higher-order terms, the two perfect matchings contribute quantum terms $\ket{0000}_{abcd}$ and $\ket{1100}_{abcd}$ with weights $\omega_{a,d}^{0,0} \omega_{b,c}^{0,0}$ and $\omega_{a,c}^{1,0} \omega_{b,d}^{1,0}$ (the product of edge weights in each matching). Their coherent superposition yields
\begin{equation}
    \ket{\psi}\approx\omega_{a,c}^{1,0} \omega_{b,d}^{1,0} \ket{1100} + \omega_{a,d}^{0,0} \omega_{b,c}^{0,0}\ket{0000}= \Big(\omega_{a,c}^{1,0} \omega_{b,d}^{1,0} \ket{11} + \omega_{a,d}^{0,0} \omega_{b,c}^{0,0}\ket{00}\Big)_{ab}\otimes \ket{00}_{cd}.
    \label{eq:Bell}
\end{equation}
Setting all weights equal and normalizing yields a two-particle Bell state between paths $a$ and $b$ that never interacted.

\section{Solutions discovered by Pytheus}

\begin{figure*}[!h]
    \centering
    \includegraphics[width =1\textwidth]{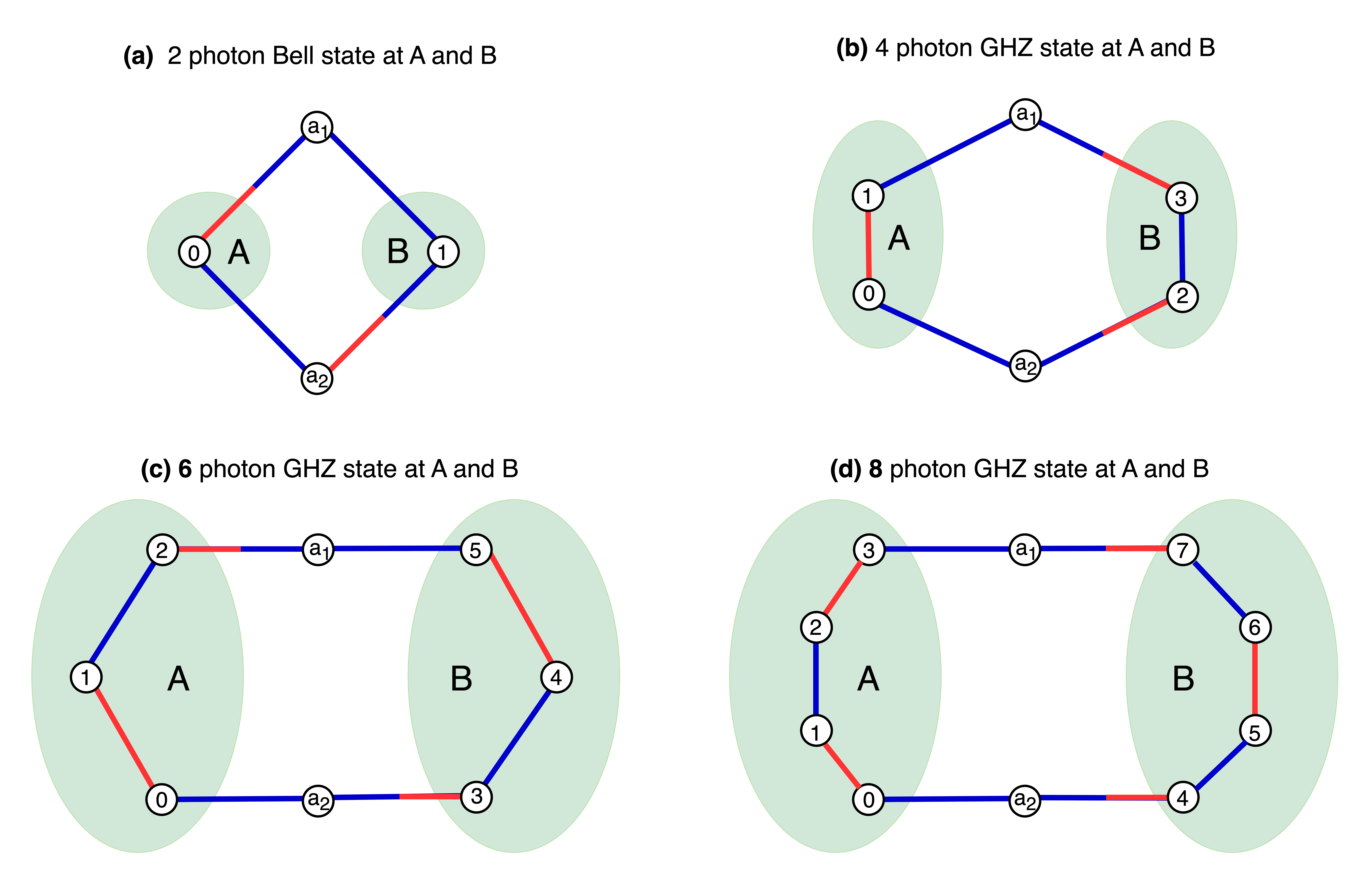}
    \caption{\textbf{Graph solutions discovered by \pytheus for generating GHZ states at two different locations with photons that never met in Fig.\ref{fig:GHZ}}. \textbf{(a)} A two-photon Bell state $\ket{\Phi^+}$, can be considered as the two-photon case of a GHZ state. \textbf{(b–d)} Four-, six-, and eight-photon GHZ states distributed across two locations. Photons at different sites that never met become entangled.}
    \label{fig:GHZgraph}
\end{figure*}

\begin{figure*}[!h]
    \centering
    \includegraphics[width =0.8\textwidth]{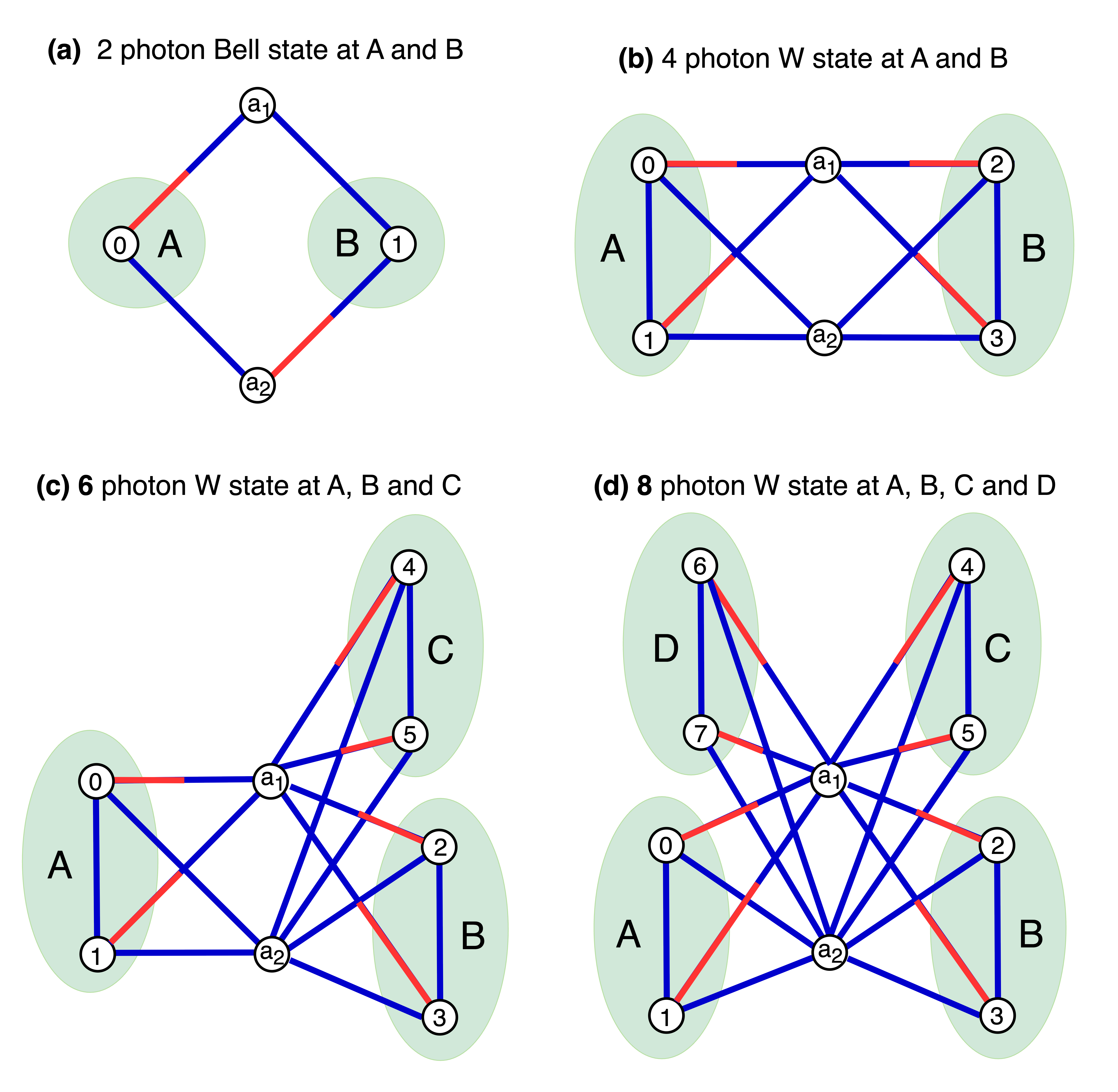}
    \caption{\textbf{Graph solutions discovered by \pytheus for generating W states at different locations with photons that never met in Fig.\ref{fig:Wstate}}. \textbf{(a)} A two-photon Bell state $\ket{\Psi^+}$, can be considered as the two-photon case of a W state. \textbf{(b–d)} Four-, six-, and eight-photon W states distributed across two, three, and four locations. Photons at different sites that never met become entangled.}
    \label{fig:wgraph}
\end{figure*}

\begin{figure}[!h]
    \centering
    \begin{subfigure}[t]{0.45\textwidth}
        \centering
        \includegraphics[width=\textwidth]{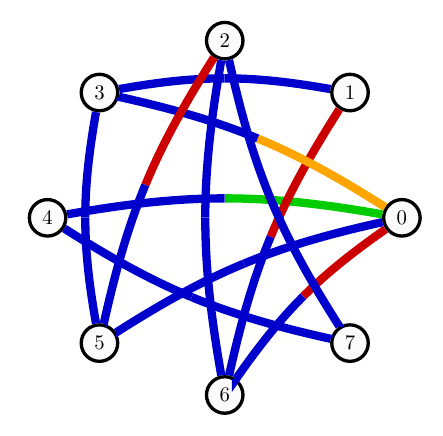}
        \caption{Graph solution for SRV(4,2,2) states at three locations discovered by \pytheus.}
        \label{fig:SRV_graph}
    \end{subfigure}
    \hfill
    \begin{subfigure}[t]{0.45\textwidth}
        \centering
        \includegraphics[width=\textwidth]{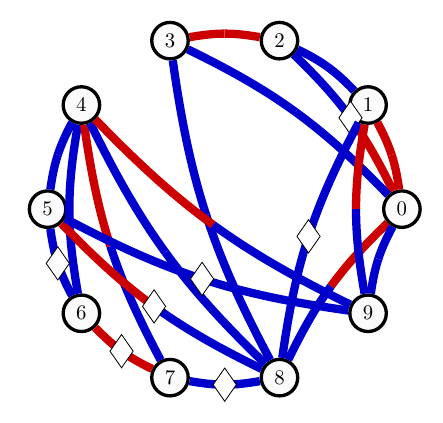}
        \caption{Graph solution for Logical Bell state with the error-detection code $[[4,1,2]]$ encoding at two locations discovered by \pytheus. }
        \label{fig:surface_graph}
    \end{subfigure}
    
    \caption{\textbf{Graph solutions discovered by \pytheus for SRV(4,2,2) state in Fig.~\ref{fig:SRV422} and logical Bell state in Fig.~\ref{fig:surface_code}.} 
    (a) SRV(4,2,2) state at three locations. Ancillary photons are at paths 3–7 (corresponding to $a_1, a_5, a_3, a_2, a_4$ in Fig.~\ref{fig:SRV422}). 
        The photons in paths 0–2, which never directly interacted, become entangled.
    (b) Logical Bell state with error-detection code $[[4,1,2]]$ encoding at two locations. Each logical qubit is encoded into four physical qubits (photons), located at sites $A$ (paths 0–3) and $B$ (paths 4–7). Ancillary photons at paths 8 and 9 (indicated as $a_1$, $a_2$ in Fig.~\ref{fig:surface_code}). The squares associated with the edges indicate negative amplitudes.}
    \label{fig:combined}
\end{figure}

\begin{figure}[t]
    \centering
    \begin{subfigure}[b]{0.45\textwidth}
        \centering
        \includegraphics[width=\textwidth]{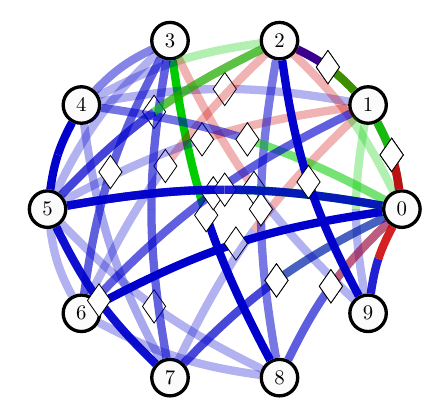}
        \caption{Hybrid high-dimensional logical–physical Bell state with the three-qutrit code $[[3,1,2]]_3$ encoding.}
        \label{fig:CSS_graph}
    \end{subfigure}
    \hfill
    \begin{subfigure}[b]{0.45\textwidth}
        \centering
        \includegraphics[width=\textwidth]{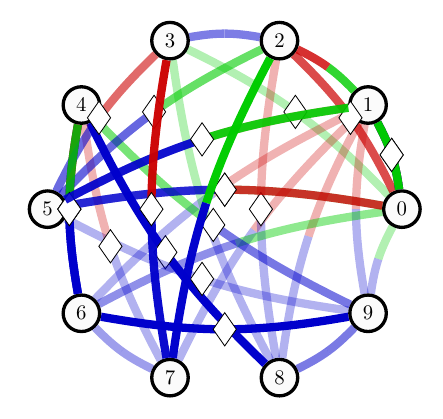}
        \caption{Hybrid high-dimensional logical–physical Bell state with the amplitude-damping code $[[4,1]]_3$ encoding.}
        \label{fig:ADcode_graph}
    \end{subfigure}
    \caption{\textbf{Graph solutions for hybrid high-dimensional logical–physical Bell states.} (a) Three-qutrit code $[[3,1,2]]_3$. Ancillary photons are in paths 4–9. The logical state is encoded in photons at paths 0-2, while the other physical qudit photon is at path 3. There is no connection between photons at location A (paths 0-2) and location B (path 3).(b) Amplitude-damping code $[[4,1]]_3$. Ancillary photons are in paths 5-9. The logical state is encoded in photons at paths 0-3, while the other physical qudit photon is at path 4. There is no connection between photons at location A (paths 0-3) and location B (path 4).}
    \label{fig:hybrid_graphs}
\end{figure}

\end{document}